\documentstyle[12pt,titlepage]{article}
\input epsfig.sty

\font\grande=cmr9.5 scaled \magstep4
\font\medio=cmr9.5 scaled \magstep2
\outer\def\beginsection#1\par{\medbreak\bigskip
      \message{#1}\leftline{\bf#1}\nobreak\medskip
\vskip-\parskip
      \noindent}

\def\laq{\raise 0.4ex\hbox{$<$}\kern -0.8em\lower 0.62
ex\hbox{$\sim$}}
\def\gaq{\raise 0.4ex\hbox{$>$}\kern -0.7em\lower 0.62
ex\hbox{$\sim$}}

\begin{document}
\bibliographystyle {unsrt}

\titlepage

\begin{flushright}
CERN-PH-TH/2008-182
\end{flushright}

\vspace{15mm}
\begin{center}
{\grande Dynamical suppression of non-adiabatic modes }\\
\vspace{15mm}
 Massimo Giovannini 
 \footnote{Electronic address: massimo.giovannini@cern.ch} \\
\vspace{6mm}

\vspace{0.3cm}
{{\sl Department of Physics, Theory Division, CERN, 1211 Geneva 23, Switzerland }}\\
\vspace{1cm}
{{\sl INFN, Section of Milan-Bicocca, 20126 Milan, Italy}}
\vspace*{2cm}

\end{center}

\vskip 2cm
\centerline{\medio  Abstract}
Recent analyses of the WMAP 5-year data constrain possible non-adiabatic contributions to the initial conditions of CMB anisotropies. Depending upon the early dynamics of the plasma, the amplitude of the entropic modes can experience a 
different suppression by the time of photon decoupling. Explicit examples of the latter observation are presented both 
analytically and numerically when 
the post-inflationary dynamics is dominated by a stiff  contribution.  
\noindent

\vspace{5mm}

\vfill
\newpage
The initial conditions of the Boltzmann hierarchy can be usefully 
classified into adiabatic and non-adiabatic (i.e. entropic).  The thermodynamic 
origin of this classification resides in the observation that the relative fluctuations 
of the specific entropy ${\mathcal S}_{\mathrm{ij}}$  do not necessarily vanish (as in the 
case of the adiabatic mode). The relative fluctuations of the specific entropy $\varsigma_{\mathrm{ij}}$ can be written, in gauge-invariant terms, as 
\begin{equation}
{\mathcal S}_{\mathrm{ij}} = \frac{\delta \varsigma_{\mathrm{ij}}}{\varsigma_{\mathrm{ij}}}
 \equiv  - 3 ( \zeta_{\mathrm{i}} - \zeta_{\mathrm{j}}), \qquad \zeta_{\mathrm{i}} = - \Psi + \frac{\delta_{\mathrm{i}}}{w_{i} +1 },
\label{ONE}
\end{equation}
where $w_{\mathrm{i}}$ is the barotropic index of the i-th species; in Eq. (\ref{ONE}) 
$\Psi$ represents the gauge-invariant Bardeen potential and $\delta_{\mathrm{i}}$ is the gauge-invariant density contrast.  In the $\Lambda$CDM model 
(where $\Lambda$ stands for the dark-energy component and CDM stands for the cold dark matter contribution) the indices i and j of Eq. (\ref{ONE}) run over the four species of the plasma so that, in general, the initial conditions will contemplate one adiabatic mode and four non-adiabatic modes (see, for instance, \cite{hannu1,hannu2}). The initial conditions 
of the Boltzmann hierarchy can be set either by choosing only the adiabatic mode, or by selecting a combination of the 
adiabatic mode with one (or more) non-adiabatic modes. The obtained 
 angular power spectra (both for temperature and polarization) can then be compared with the experimental data and interesting bounds can be set on the various combinations of the initial conditions\footnote{According to the present data, the adiabatic mode will have to be dominant in comparison with the remaining one (or more) entropic 
contributions. In the opposite case the (observed) anti-correlation peak in the temperature/polarization would not be correctly reproduced.} \cite{WMAP51,WMAP52} (see also \cite{hannu1,hannu2}). 

Are there simple dynamical recipes able to suppress the entropic 
contributions? This is the basic question addressed in this paper.
In the current framework, after inflation, the plasma was suddenly dominated by radiation. Absent the latter assumption, the  non-adiabatic contribution  to the pre-decoupling fluctuations of the spatial curvature will have a
different relation to the entropic modes originally present right after inflation.  
While it is not mandatory to postulate different dynamical evolutions, 
it is useful to be aware of different possibilities which can help more dedicated scrutiny of the observational data. 

Prior to the radiation epoch, the plasma might have been expanding 
at a slower rate. This perspective was also invoked by Zeldovich who 
suggested that, prior to radiation dominance, the Universe 
was indeed quite stifff and characterized by a sound speed even comparable with the speed of light \cite{zel1}. Post-inflationary phases stiffer 
than radiation can even lead to relic gravitons whose spectral energy density increases as a function of the comoving frequency \cite{max1}.  
If the inflaton field is identified with the quintessence field a stiff post-inflationary phase arises naturally \cite{max2} and this is what happens in the context of the so-called quintessential inflationary models \cite{PV} 
as well as in related contexts  \cite{kessence1,kessence2}. 

Consider, for sake of simplicity, a post-inflationary plasma 
characterized by three distinct  components so that the total energy density 
and the various pressures can be written as:
\begin{equation}
\rho_{\mathrm{t}} = \rho_{\mathrm{m}} +  \rho_{\mathrm{r}} + \rho_{\mathrm{S}},\qquad 
p_{\mathrm{m}} =0,\qquad p_{\mathrm{r}} = \frac{\rho_{\mathrm{r}}}{3}\qquad p_{\mathrm{S}} = w \rho_{\mathrm{S}},
\label{EX2a}
\end{equation}
where $\rho_{\mathrm{r}}$ and $\rho_{\mathrm{m}}$ denote, respectively, the radiation and the matter energy densities while $p_{\mathrm{r}}$ and $p_{\mathrm{m}}$  indicate the corresponding pressures. 
In Eq. (\ref{EX2a}) $\rho_{\mathrm{S}}$  and $p_{\mathrm{S}}$ are the energy density and pressure of a supplementary component whose generic barotropic index will simply be denoted by $w$.  From the pertinent 
Friedmann-Lema\^itre equations \footnote{A conformally flat metric 
$g_{\mu\nu} =a^2(\tau) \eta_{\mu\nu}$ will be assumed throughtout. The prime denotes a derivation 
with respect to the conformal time coordinate $\tau$.}
\begin{equation}
{\mathcal H}^2 = \frac{8 \pi G a^2}{3} \rho_{\mathrm{t}},\qquad {\mathcal H}^2 - {\mathcal H}' = 4 \pi Ga^2 (p_{\mathrm{t}} + \rho_{\mathrm{t}}), \qquad {\mathcal H} = \frac{a'}{a},
\label{FL}
\end{equation}
it can be easily  argued that, for $w> 1/3$, $\rho_{\mathrm{S}}$ dominates (at 
early times) in comparison with $\rho_{\mathrm{m}}$ and $\rho_{\mathrm{r}}$.
In Eqs. (\ref{FL}), $\rho_{\mathrm{S}} \simeq a^{-3(w + 1)}$ while 
$\rho_{\mathrm{r}} \simeq a^{- 4}$; ergo $\rho_{\mathrm{S}}$ will decrease 
faster than $\rho_{\mathrm{r}}$ and this is the reason why the backreaction 
of the radiation can be important \cite{max1,max2,PV}. 

For the present purposes, it is practical to separate 
the adiabatic and the entropic fluctuations composing the gauge-invariant 
perturbation of the total pressure i.e. 
\begin{eqnarray}
&& \delta p_{\mathrm{t}} = c_{\mathrm{st}}^2 \delta \rho_{\mathrm{t}} + \delta p_{\mathrm{nad}}, 
\label{deltap}\\
&& c_{\mathrm{st}}^2 = \biggl(\frac{\delta p_{\mathrm{t}}}{\delta \rho_{\mathrm{t}}}\biggr)_{\varsigma_{\mathrm{ij}}} = \sum_{\mathrm{i
}} \frac{\rho_{\mathrm{i}}'}{\rho_{\mathrm{t}}'} c_{\mathrm{si}}^2, \qquad 
\delta p_{\mathrm{nad}} =  \biggl(\frac{\delta p_{\mathrm{t}}}{\delta \varsigma_{\mathrm{ij}}}\biggr)_{\rho_{\mathrm{t}}}  \delta \varsigma_{\mathrm{ij}},
\label{defdelta}
\end{eqnarray}
where $\delta\varsigma_{\mathrm{ij}}$ are the fluctuations in the specific entropy already introduced in Eq. (\ref{ONE}).
In Eq. (\ref{defdelta})  the subscripts in the round brackets remind that the variation must be taken, respectively, 
for $\delta \varsigma_{\mathrm{ij}} =0$ and for $\delta \rho_{\mathrm{t}} =0$.
Since, according to Eq. (\ref{EX2a}), the plasma is composed by three 
species there will be, in general, three entropic contributions.  
Equation (\ref{defdelta}) allows, indeed, for a more explicit form of $\delta p_{\mathrm{nad}}$:
\begin{equation}
\delta p_{\mathrm{nad}} = \frac{1}{6 {\mathcal H} \rho_{\mathrm{t}}'} \sum_{\mathrm{i\,j}} \rho_{\mathrm{i}}' \rho_{\mathrm{j}}' (c_{\mathrm{si}}^2 - c_{\mathrm{sj}}^2) {\mathcal S}_{\mathrm{ij}}, \qquad c_{\mathrm{si}}^2 = \frac{p_{\mathrm{i}}'}{\rho_{\mathrm{i}}'},
\label{dpnad}
\end{equation}
where the summation indices run over the three (or more) species of the fluid mixture.
Equation (\ref{dpnad})  directly follows from Eqs. (\ref{deltap}) and (\ref{defdelta}) by considering 
a generic pair of fluids and by summing over all the components. 
Using Eq. (\ref{EX2a}) into Eq. (\ref{dpnad}), $\delta p_{\mathrm{nad}}$ can be explicitly 
obtained:
\begin{equation}
\delta p_{\mathrm{nad}} = - \frac{1}{9{\mathcal H} \rho_{\mathrm{t}}'} [ \rho_{\mathrm{m}}' \rho_{\mathrm{r}}' {\mathcal S}_{\mathrm{mr}} + 3 w \rho_{\mathrm{m}}' \rho_{\mathrm{S}}' {\mathcal S}_{\mathrm{mS}} - ( 3 w - 1) \rho_{\mathrm{S}}' \rho_{\mathrm{r}}' {\mathcal S}_{\mathrm{Sr}}],
\label{EX3a}
\end{equation}
where $ {\mathcal S}_{\mathrm{mr}}$, ${\mathcal S}_{\mathrm{mS}}$ and ${\mathcal S}_{\mathrm{Sr}}$ are, respectively, the three independent entropic fluctuations which can arise in the problem. 
To pass correctly from Eq. (\ref{dpnad}) to (\ref{EX3a}) it should be borne in mind that 
${\mathcal S}_{\mathrm{ij}} = - {\mathcal S}_{\mathrm{ji}}$.  In a democratic perspective 
all the entropic contributions in Eq. (\ref{EX3a})  can be present and with comparable amplitude. In the complementary  situation one of the terms (e. g. ${\mathcal S}_{\mathrm{mr}}$) is much larger than the remaining two. 

\begin{figure}
\begin{center}
\begin{tabular}{|c|c|}
      \hline
      \hbox{\epsfxsize = 7.7 cm  \epsffile{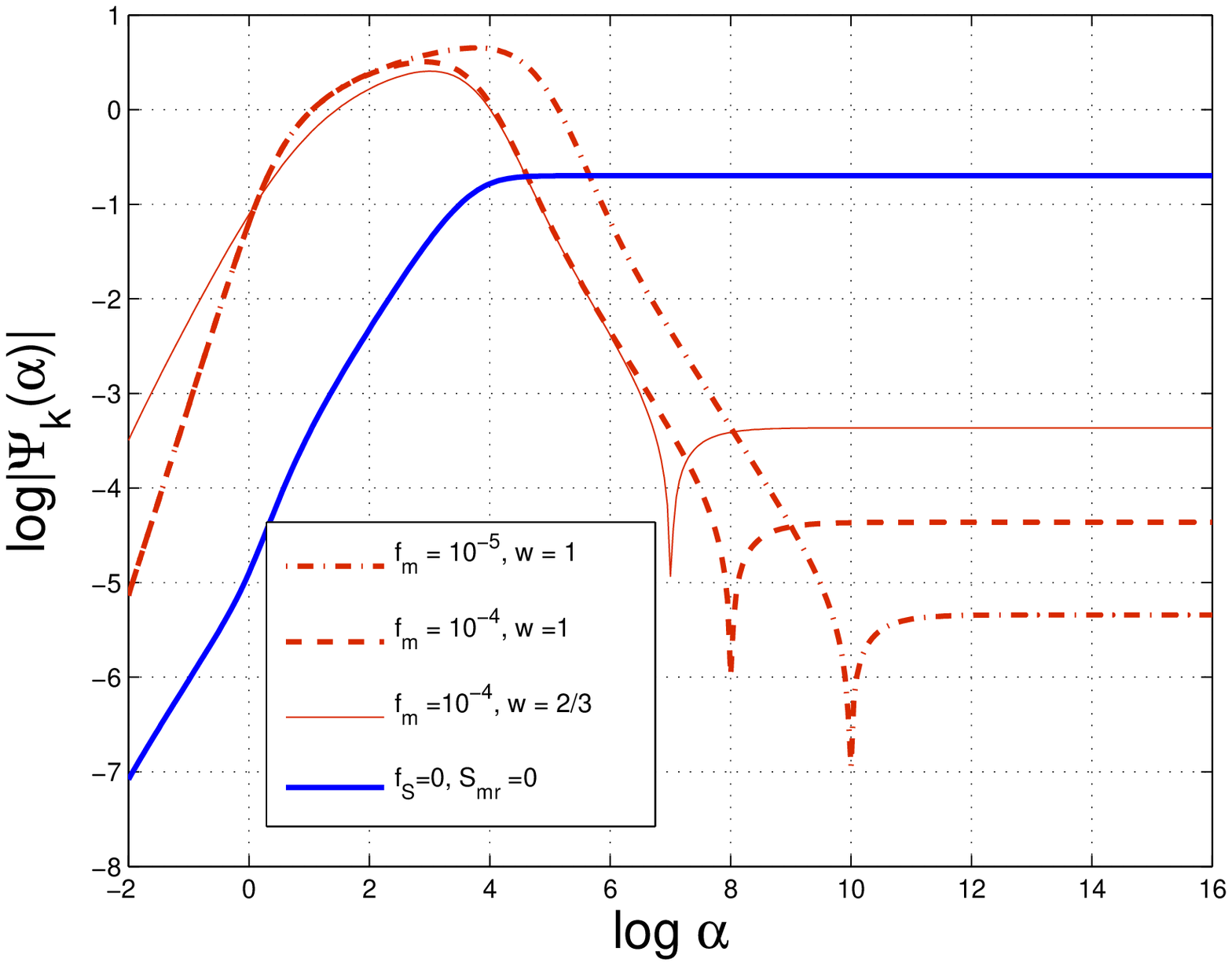}} &
      \hbox{\epsfxsize = 7.7 cm  \epsffile{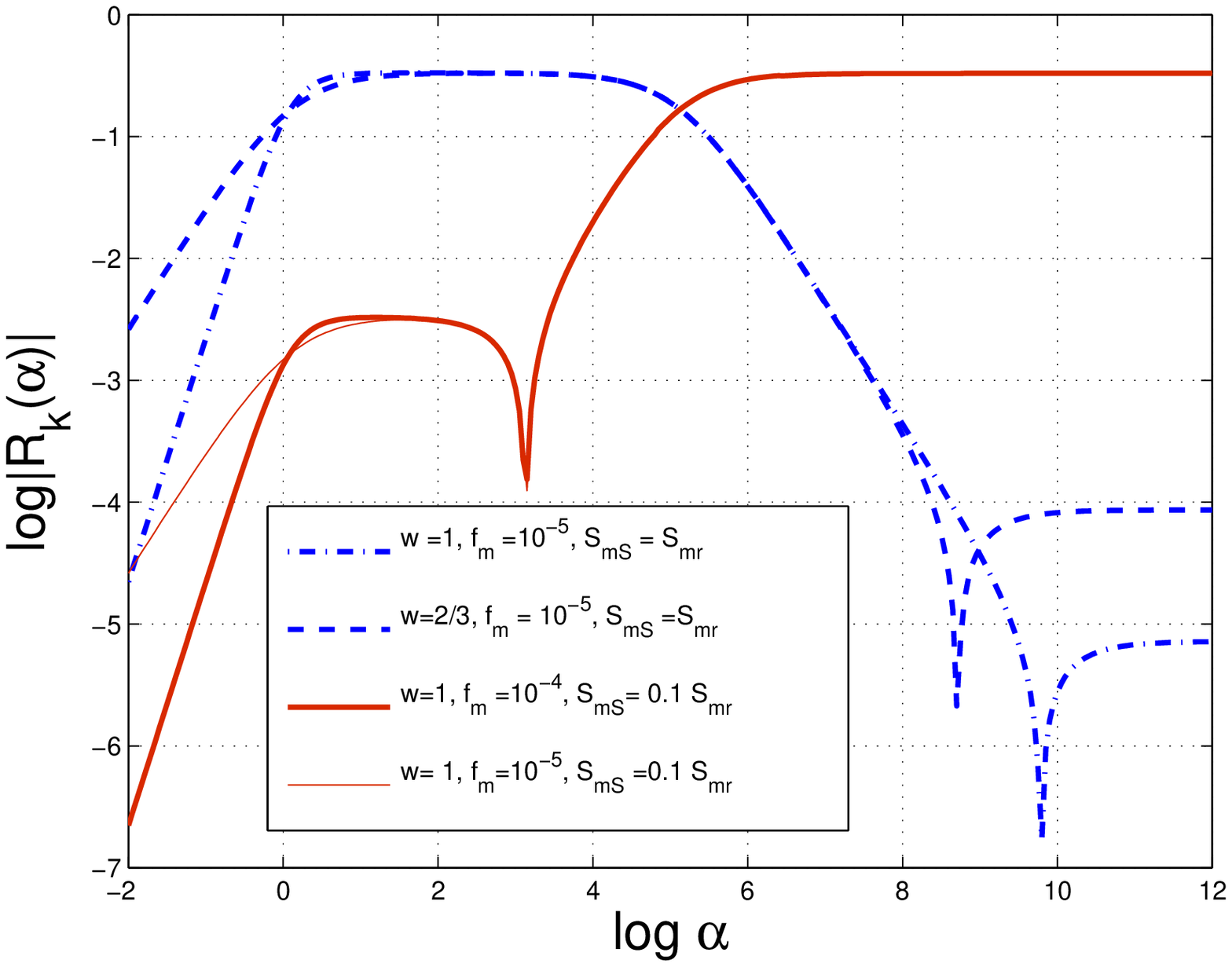}}\\
      \hline
\end{tabular}
\end{center}
\caption[a]{The evolutions of $\Psi_{k}$  (plot at the left) and of ${\mathcal R}_{k}$ 
(plot at the right) are illustrated. In the right plot the cases  ${\mathcal S}_{\mathrm{mS}}= {\mathcal S}_{\mathrm{mr}}$ and ${\mathcal S}_{\mathrm{mS}}\ll {\mathcal S}_{\mathrm{mr}}$ are compared. }
\label{F2}
\end{figure}

The fate of the non-adiabatic contributions given in Eq. (\ref{EX3a}) can be 
determined from the gauge-invariant evolution equations of the curvature and metric inhomogeneities. The gauge-invariant form of the Hamiltonian and the momentum constraints is
\begin{eqnarray} 
&&\nabla^2 \Psi = 4\pi G a^2 \rho_{\mathrm{t}} \epsilon_{\mathrm{t}}, \qquad 
 \nabla^2 ( {\mathcal H} \Phi + \Psi') = - 4\pi G a^2 (p_{\mathrm{t}} + \rho_{\mathrm{t}}) \theta_{\mathrm{t}},
\label{hammom}\\
&& (p_{\mathrm{t}} + \rho_{\mathrm{t}}) \theta_{\mathrm{t}}  = \sum_{\mathrm{i}} (p_{\mathrm{i}} + \rho_{\mathrm{i}}) \theta_{\mathrm{i}},
\label{TOTVEL}
\end{eqnarray}
where $\theta_{\mathrm{t}}$ is the three-divergence of the (total) velocity 
field (defined in (\ref{TOTVEL}) as the sum of the velocities of the individual fluids) 
and $\epsilon_{\mathrm{t}}$ is the gauge-invariant density contrast which corresponds to the total density contrasts in the comoving orthogonal gauge \cite{hoff,max3,KS}. 
In terms of $\epsilon_{\mathrm{t}}$ the momentum constraint (i.e. first relation 
of Eq. (\ref{hammom})) takes a form which is reminiscent of the (non-relativistic) 
Poisson equation. It turns out that $\epsilon_{\mathrm{t}}$ is proportional 
to the difference of other two useful gauge-invariant quantities:
\begin{equation}
 \epsilon_{\mathrm{t}} = \frac{3 ( \rho_{\mathrm{t}} + p_{\mathrm{t}})}{\rho_{\mathrm{t}}} (\zeta - {\mathcal R}),
\label{eps1}
\end{equation}
where 
\begin{equation}
\zeta = \sum_{\mathrm{i}} \frac{\rho_{\mathrm{i}}'}{\rho_{\mathrm{t}}'} \zeta_{\mathrm{i}} \equiv - \Psi - \frac{\delta\rho_{\mathrm{t}} {\mathcal H}}{\rho_{\mathrm{t}}'},\qquad 
 {\mathcal R} = - \Psi - \frac{{\mathcal H}({\mathcal H} \Phi + \Psi')}{4 \pi G a^2 (p_{\mathrm{t}} + \rho_{\mathrm{t}})}.
\label{zeta1R1}
\end{equation}
The gauge-invariant variable $\zeta$ can be interpreted either as the 
curvature perturbation in the uniform density gauge or as 
the density contrast in the uniform curvature gauge 
\cite{hoff,max3,KS} (see also \cite{noh,hwang}). The variable ${\mathcal R}$ represent the (gauge-invariant) curvature perturbations which effectively correspond to the fluctuations of the spatial curvature on comoving orthogonal hypersurfaces. The evolution of $\zeta$ can be easily obtained from the equation for the total density fluctuation derived from the perturbation of the covariant conservation of the (total) energy-momentum tensor, i.e.  
\begin{equation}
\delta \rho_{\mathrm{t}}' - 3 \Psi' (p_{\mathrm{t}} + \rho_{\mathrm{t}})  + (p_{\mathrm{t}} + \rho_{\mathrm{t}}) \theta_{\mathrm{t}} + 
3 {\mathcal H} ( \delta p_{\mathrm{t}}+\delta \rho_{\mathrm{t}}) =0.
\label{delta1}
\end{equation}
From Eq. (\ref{zeta1R1}) it can be easily deduced that $\delta \rho_{\mathrm{t}} = 3 (p_{\mathrm{t}} + \rho_{\mathrm{t}}) ( \zeta + \Psi)$. Using 
the latter relation inside Eq. (\ref{delta1}) the evolution of $\zeta$ is  simply
\begin{equation}
\zeta' = - 
\frac{{\mathcal H} }{p_{\mathrm{t}} + \rho_{\mathrm{t}}}\,\delta p_{\mathrm{nad}} - \frac{\theta_{\mathrm{t}}}{3},
\label{zeta2}
\end{equation}
where Eqs. (\ref{FL}) and (\ref{deltap}) have been used. 
Recalling Eqs. (\ref{eps1})--(\ref{zeta1R1}) the evolution equation for ${\mathcal R}$ 
can be directly obtained and it is
\begin{equation}
{\mathcal R}' = - \frac{{\mathcal H}}{ p_{\mathrm{t}} +\rho_{\mathrm{t}}}\,\delta p_{\mathrm{nad}} + \frac{{\mathcal H}}{12 \pi G a^2 (p_{\mathrm{t}} + \rho_{\mathrm{t}})} \nabla^2 (\Phi - \Psi) - \frac{{\mathcal H} c_{\mathrm{st}}^2}{4\pi G a^2 (p_{\mathrm{t}} + \rho_{\mathrm{t}})} \nabla^2 \Psi.
\label{R2}
\end{equation}
\begin{figure}
\begin{center}
\begin{tabular}{|c|c|}
      \hline
      \hbox{\epsfxsize = 7.7 cm  \epsffile{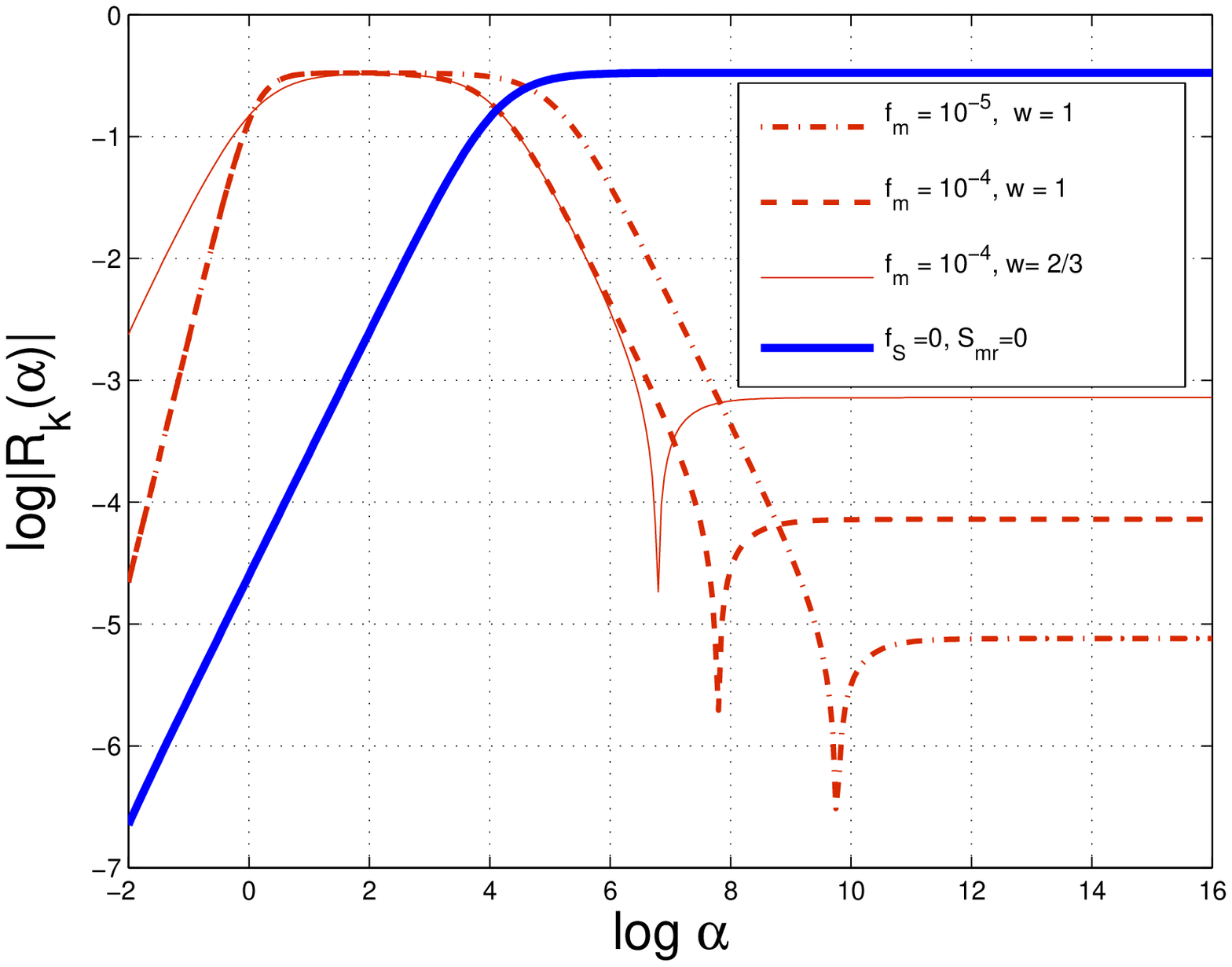}} &
      \hbox{\epsfxsize = 7.7 cm  \epsffile{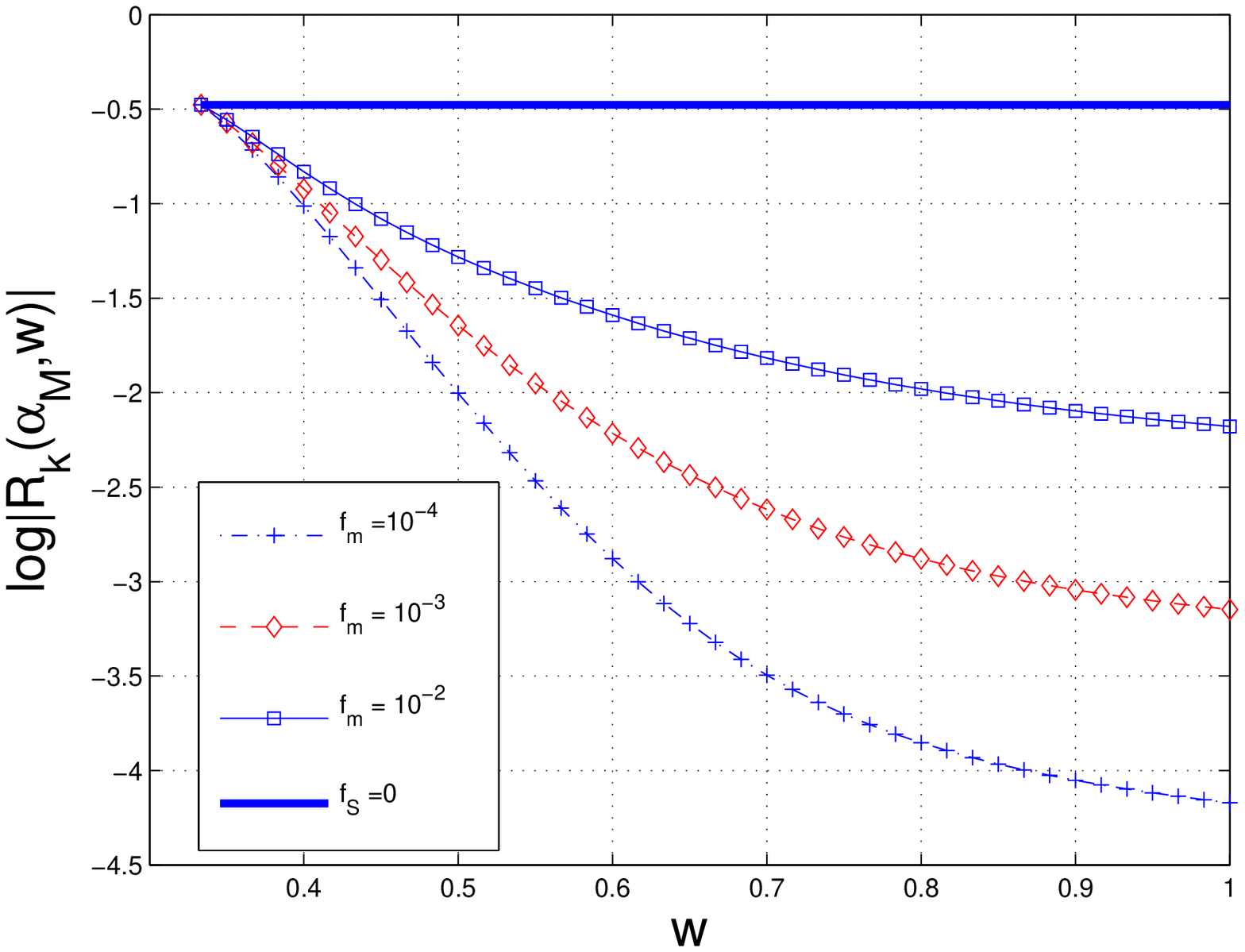}}\\
      \hline
\end{tabular}
\end{center}
\caption[a]{The evolution of the curvature perturbations as a function of the scale factor and for various fixed 
values of $w$ (plot at the left); the evolution of curvature perturbations for fixed $\alpha\gg 1$ and as a function of the sound speed of the stiff component (plot at the right).}
\label{F1}
\end{figure}
The dynamical content of Eqs. (\ref{R2}) and Eq. (\ref{zeta2}) is clearly the same, 
i.e. the two equations coincide exactly in the long wavelength limit (i.e. $k\tau\ll 1$).
Since $\rho_{\mathrm{S}} \neq 0$ in Eq. (\ref{EX2a}), 
from Eqs. (\ref{EX3a}) and (\ref{R2}) the evolution for the curvature perturbations becomes then, to leading order in $k\tau$, 
\begin{equation}
{\mathcal R}_{k}'  = {\mathcal F}_{{\mathcal R}}(\tau, w, f_{\mathrm{S}}, f_{\mathrm{M}}), \qquad \Psi_{k}'  = - {\mathcal G}_{1}(\tau, w, f_{\mathrm{S}}, f_{\mathrm{M}}) \Psi_{k} - 
{\mathcal G}_{2}(\tau, w, f_{\mathrm{S}}, f_{\mathrm{M}}) {\mathcal R}_{k}.
\label{Rex1}
\end{equation}
In Eq. (\ref{Rex1}) the function ${\mathcal F}_{{\mathcal R}}(\tau, w, f_{\mathrm{S}}, f_{\mathrm{M}})$ is given by 
\begin{equation}
 \frac{{\mathcal H} \{ ( w + 1)[ 4  ( 3 w -1) f_{\mathrm{S}} {\mathcal S}_{\mathrm{rS}} \alpha^{3 w -1} - 9 w f_{\mathrm{m}} f_{\mathrm{S}} {\mathcal S}_{\mathrm{mS}} \alpha^{3 w}] - 4 f_{\mathrm{m}} {\mathcal S}_{\mathrm{mr}} \alpha^{6w -1}  \}}{[4 \alpha^{3 w-1} + 3\alpha^{3 w} f_{\mathrm{m}} + 3 (w + 1) f_{\mathrm{S}}]^2},
\label{EX4a}
\end{equation}
while the remaining two functions are:
\begin{eqnarray}
&& {\mathcal G}_{1}(\tau, w, f_{\mathrm{S}}, f_{\mathrm{M}})= {\mathcal H}\frac{[6 \alpha^{w-1} + (3 w + 5) f_{\mathrm{S}} + 5 f_{\mathrm{m}} \alpha^{3 w}]}{2 \alpha [ \alpha^{3 w-1} + 
f_{\mathrm{S}} + f_{\mathrm{m}} \alpha^{3 w}]},
\label{EX4b}\\
&& {\mathcal G}_{2}(\tau, w, f_{\mathrm{S}}, f_{\mathrm{M}})={\mathcal H}\frac{[3 f_{\mathrm{m}} \alpha^{3 w} + 4 \alpha^{3 w -1} + 3 (w + 1) f_{\mathrm{S}}]}{2 \alpha [ \alpha^{3 w-1} + 
f_{\mathrm{S}} + f_{\mathrm{m}} \alpha^{3 w}]},
\label{EX4c}
\end{eqnarray}
having introduced the following rescalings: 
\begin{equation}
\alpha = \frac{a}{a_{*}}, \qquad f_{\mathrm{m}} = \frac{\rho_{\mathrm{m}}(a_{*})}{\rho_{\mathrm{r}}(a_{*})}, \qquad 
f_{\mathrm{S}} = \frac{\rho_{\mathrm{S}}(a_{*})}{\rho_{\mathrm{r}}(a_{*})},\qquad 
a_{*} = a(\tau_{*}).
\label{EX5a}
\end{equation}
  In what follows we will always take $f_{\mathrm{S}}=1$ and set initial conditions for the numerical integration 
for $\alpha_{\mathrm{i}} = a_{\mathrm{i}}/a_{*} \ll 1$.  It is worth stressing that the condition $f_{\mathrm{S}} =1$ 
implies that $\rho_{\mathrm{S}}(a_{*}) = \rho_{\mathrm{r}}(a_{*})$. The value of $f_{\mathrm{m}}< 1$
 specifies the fraction of non-relativistic matter eventually present for $\tau_{*}$. 
The results of the numerical integration are reported in Figs. \ref{F1} and \ref{F2} for some illustrative sets 
of parameters.
 With the full line (plot at the left in Fig. \ref{F1})we have the result for the case ${\mathcal S}_{\mathrm{rS}} = {\mathcal S}_{\mathrm{mS}} =0$. In this 
 case it is well known that the non-adiabatic solution to $|{\mathcal R}_{k}(\alpha)|$ goes as ${\mathcal S}_{\mathrm{mr}}/3$ (units ${\mathcal S}_{\mathrm{mr}}= 1$ will be adopted throughout). This is exactly the result 
 of the full line reported in Fig. \ref{F1} (plot at the left)  where, asyptotically for $\alpha \gg1$,   $\log{|{\mathcal R}_{k}|} \simeq -0.477$.  In the limit $f_{\mathrm{S}}\to 0$, the equation for ${\mathcal R}_{k}$ (see  Eq. (\ref{Rex1}))
can be written as 
\begin{equation}
{\mathcal R}_{k}' = - 4 \frac{f_{\mathrm{m}}{\mathcal S}_{\mathrm{mr}}}{(4 + 3 f_{\mathrm{m}}\alpha)} + {\mathcal O}(k^2\tau^2), \qquad 
{\mathcal R}_{k}(\tau) \simeq {\mathcal R}_{*}(k) - \frac{{\mathcal S}_{\mathrm{mr}}(k)}{3}+ {\mathcal O}(k^2 \tau^2),
\label{FIVE}
\end{equation}
where  ${\mathcal R}_{*}(k)$ parametrizes the adiabatic solution which has been 
added for completeness but which will be left untouched by the present considerations. 
Equation (\ref{FIVE}) gives the value of the curvature perturbations induced by the non-adiabatic CDM-radiation mode when, right after inflation the Universe is dominated by a radiative equation of state.
Recalling the form of (ordinary)  Sachs-Wolfe contribution to the temperature 
anisotropies we will than have that, in the sudden decoupling approximation, 
$\Delta_{\mathrm{T}}^{\mathrm{(SW)}} \simeq  
- {\mathcal R}_{*}(k)/5  + 2 {\mathcal S}_{*}(k)/5$. If the 
fluids do not exchange energy and momentum, as customarily assumed 
in the simplest situation and as verified in our case, then,  ${\mathcal S}_{ij}'=0$ in the long wavelength limit. 
Indeed, it can be easily shown that $\delta_{\mathrm{i}}' = 3(w_{\mathrm{i}} +1) \Psi' - 
(w_{\mathrm{i}} + 1) \theta_{\mathrm{i}}$ where $\theta_{\mathrm{i}} = \vec{\nabla}\cdot 
\vec{v}_{\mathrm{i}}$. Equation (\ref{ONE}) then implies ${\mathcal S}_{\mathrm{ij}}' = 
(\theta_{\mathrm{i}} - \theta_{\mathrm{j}})$, i.e. ${\mathcal S}_{\mathrm{ij}}'=0$ 
up to corrections ${\mathcal O}(k^2\tau^2)$. 

When $\rho_{\mathrm{S}}\neq 0$, in the generic situation all the ${\mathcal S}_{\mathrm{ij}}$ are non-vanishing. Absent any specific knowledge 
of the initial conditions, the various entropic fluctuations 
can be expected to be comparable, i.e. ${\mathcal S}_{\mathrm{rS}}(k) \simeq {\mathcal S}_{\mathrm{mS}}(k) \simeq {\mathcal S}_{\mathrm{mR}}$.
If we take, for instance, $f_{\mathrm{m}} = 10^{-2}$ (dashed line in the left plot) 
${\mathcal R}_{k}(\alpha)$ gets first to $1/3$ (as implied by Eq. (\ref{FIVE}) 
in the absence of $\rho_{\mathrm{S}}$ ) and then decreases by reaching, subsequently, 
a constant value which is between two and three orders 
of magnitude smaller than the putative asymptotic value which characterizes 
the case ${\mathcal S}_{\mathrm{rS}} = {\mathcal S}_{\mathrm{mS}} =0$ (i.e. $1/3$). 
As $f_{\mathrm{m}}$ decreases the intermediate plateau get larger, 
and the asymptotic value gets progressively reduced.  In the left plot of Fig. \ref{F1}
it has been assumed that $w = 1$. In the plot at the right ${\mathcal R}_{k}(\alpha)$ 
is reported for a fixed value of $\alpha$ (i.e., more specifically, $\alpha_{\mathrm{M}} = 10^{9}$) but as a function of $w$. As $w\to 1$ the
 suppression can even be, depending on the parameters ${\mathcal O}(10^{-5})$. Of course the specific figure depends upon the other parameters. At  the same 
 time it is clear that the amount of suppression depends upon the degree 
 of stiffness, i.e. upon $|w-1/3|$.

The occurence that, for a while, ${\mathcal R}_{k} \simeq {\mathcal O}(1/3)$  in Figs. \ref{F1} and \ref{F2}  just means that, depending 
upon $f_{\mathrm{m}}$ the terms containing $f_{\mathrm{S}}$ can be 
neglected for intermediate values of $\alpha$. 
The decrease in ${\mathcal R}_{k}(\alpha)$ can be also explained. The analytical estimate can be separated into two steps, i.e. between $\alpha_{\mathrm{i}}$ and $\alpha_{X} > 1$ and between 
$\alpha_{X}$ and $\alpha_{\mathrm{M}} \gg 1 $. In the first transition the terms 
proportional to $f_{\mathrm{m}}$ can be neglected. The approximate 
result will then be:
\begin{equation}
{\mathcal R}_{k}(\alpha_{X}) \simeq {\mathcal R}_{*}(k) + \biggl(\frac{4 {\mathcal S}_{\mathrm{Sr}}}{3 f_{\mathrm{S}}} \biggr) \frac{\alpha_{X}^{3 w -1}}{4 \alpha_{X}^{ 3 w -1} + 3 ( w+ 1) f_{\mathrm{S}}} \simeq \biggl(\frac{ {\mathcal S}_{\mathrm{Sr}}}{3 f_{\mathrm{S}}} \biggr) + {\mathcal O}(\alpha_{\mathrm{X}}^{-1}).
\label{ES1}
\end{equation}
For the second transition the relevant term will be the one proportional to ${\mathcal S}_{\mathrm{mr}}$ since the other terms can be neglected for $\alpha> 1$ and for $w>1/3$.
By fixing the integration constant from Eq. (\ref{ES1}) the asymptotic result (valid in the 
limit $\alpha \gg 1$) 
\begin{equation}
{\mathcal R}_{k}(\alpha_{\mathrm{M}}) \simeq {\mathcal R}_{*}(k) + \frac{{\mathcal S}_{\mathrm{Sr}}}{3f_{\mathrm{S}}} - \frac{{\mathcal S}_{\mathrm{mr}}f_{\mathrm{m}}\alpha_{\mathrm{M}}}{
3 f_{\mathrm{m}} \alpha_{\mathrm{M}} + 4} \simeq \biggl(
\frac{{\mathcal S}_{\mathrm{Sr}}}{3 f_{\mathrm{S}}} - \frac{{\mathcal S}_{\mathrm{mr}}}{3} \biggr) + \frac{ 4 {\mathcal S}_{\mathrm{mr}}}{9 \alpha_{\mathrm{M}} f_{\mathrm{m}}}.
\label{ES2}
\end{equation}
If we do not assume a large hierarchy between ${\mathcal S}_{\mathrm{Sr}}$ and 
${\mathcal S}_{\mathrm{mr}}$ the first term in the second equality of Eq. (\ref{ES2}) 
will approximately vanish and the second term will lead to a decrease of ${\mathcal R}_{k}(\alpha)$. This is the effect observed in Fig. \ref{F1}.  If the values 
of the entropic fluctuations are drastically different (e.g. ${\mathcal S}_{\mathrm{mr}}$ 
dominates against the others) then the results of Eq. (\ref{FIVE}) will be approximately 
true, at least asymptotically. This aspect is illustrated in Fig. \ref{F2} (plot at the right)
for few cases where ${\mathcal S}_{\mathrm{mr}}$ is ten times larger than the other 
entropic contributions. Always in Fig. \ref{F2} (plot at the left) the evolution of 
$\Psi_{k}$ is reported. The full line in Fig. \ref{F2} (plot at the left) 
corresponds to the case $\rho_{\mathrm{S}} =0$ where, following the same 
considerations of Eq. (\ref{FIVE}), $\Psi_{k}(\alpha) \simeq {\mathcal S}_{\mathrm{mr}}/5$, i.e. $\log{|\Psi_{k}(\alpha)|} \simeq -0.698$ in units ${\mathcal S}_{\mathrm{mr}}=1$.
The same patterns of suppressions discussed in the case of ${\mathcal R}_{k}$ also arise, as expected for $\Psi_{k}$. Note that, finally, once ${\mathcal R}_{k}$ and $\Psi_{k}$ are known to a given order in $k\tau$, the constraints of Eqs. (\ref{hammom})--(\ref{TOTVEL}) and (\ref{eps1}) can be used to derive $\epsilon_{\mathrm{t}}$, $\zeta$ and $\theta_{\mathrm{t}}$. 

The examples presented in this paper suggest that the entropic fluctuations 
can be dynamically suppressed if, after inflation, there is a stiff contribution 
to the primeval plasma. The rationale for the obtained result depends both on a 
modification of the dynamics (at early time the stiff contribution dominates) 
and upon an interference effect between the various entropic contributions. 
In fine, the exercise presented here suggests that 
 that the bounds on the non-adiabatic contribution obtained from the CMB data analysis do depend upon a number of specific assumptions on the early thermal history of the background geometry.  In different terms, 
the entropy fluctuations used to set initial conditions of the Boltzmann hierarchy
prior to equality might be already suppressed as a consequence 
of the preceding dynamical evolution.

\end{document}